# Oersted Lecture 2013:
# How should we think about how our students think?

Edward F. Redish

*Department of Physics, University of Maryland, College Park, MD 20742-4111*

**Abstract.** Physics Education Research (PER) applies a scientific approach to the question, "How do our students think about and learn physics?" PER allows us to explore such intellectually engaging questions as, "What does it mean to *understand* something in physics?" and, "What skills and competencies do we want our students to learn from our physics classes?" To address questions like these, we need to do more than observe student difficulties and build curricula. We need a theoretical framework – a structure for talking about, making sense of, and modeling how one thinks about, learns, and understands physics. In this paper, I outline some aspects of the *Resources Framework*, a structure that some of us are using to create a phenomenology of physics learning that ties closely to modern developments in neuroscience, psychology, and linguistics. As an example of how this framework gives new insights, I discuss *epistemological framing* — the role of students' perceptions of the nature of the knowledge they are learning and what knowledge is appropriate to bring to bear on a given task. I discuss how this foothold idea fits into our theoretical framework, show some classroom data on how it plays out in the classroom, and give some examples of how my awareness of the resources framework influences my approach to teaching.

**Keywords:** Physics education research, expectations, selective attention, transfer, framing, theoretical framework, cognitive modeling, socio-cultural, Oersted, epistemology, resources

**PACS:** 01.40.Fk Research in physics education, 01.40.Ha Learning theory and science teaching, 01.40.gb Teaching methods and strategies.

## I. PREAMBLE

After 20 years as a physics education researcher, I am honored to be added to the distinguished list of Oersted medalists, especially since so many of them have inspired my thinking over many years. My work over these decades has also been and continues to be facilitated, guided, and enriched by many other mentors, colleagues, and students, too many to name here.

In this paper, I want to convince you of two ideas that have emerged from and motivated my research for the past two decades:

(1) There is value in having a theoretical framework for physics education research.

(2) One value of such a framework is learning to appreciate the importance of mental control structures such as expectations, framing, and selective attention in making sense of what might first appear as students' lack of knowledge.

The paper is organized as follows: In section 2, I discuss what I mean by a theoretical framework by giving three examples from our familiar sciences. In section 3, I motivate why physics education research (PER) needs a theoretical framework based in the psychological sciences. In section 4, I walk you through a few simple experiments that you can do yourself to demonstrate some of the fundamental principles on which the framework is built. In section 5, I introduce some of the basic principles of our Resources Framework. In section 6, I discuss some examples of how these results play out in a classroom; and, in section 7, I discuss some implications for instruction and research.

## II. WHAT'S A THEORETICAL FRAMEWORK?

A growing number of scientists have been turning their attention to studying how their students learn a scientific discipline – and how they don't. These researchers want to figure out how to do a more effective job teaching their students, and to do that, they need to better understand what it means to understand science. This requires a deep understanding of the science, so this kind of research is often carried out by disciplinary scientists, and their research field is referred to as *discipline-based educational research* (DBER).[1] Physics education research (PER) is perhaps the oldest and (presently) the best established of the DBER disciplines.[2,3] AAPT's Physics

Education Research (PER) Topical Group currently has more than 750 members – physicists interested in the use of a scientific approach to study the teaching and learning of physics, and the APS has just recently created their own PER topical group.

There are typically three complementary approaches that comprise any science: observation (experiment), practice (engineering), and mechanism (theory). They intertwine, inform each other, and provide mutual support. They are like the legs of a three-legged stool, and, as we all know well, the most important leg of a three-legged stool is the one that's missing. In PER we have a tendency to focus on observation and practice: How do we see our students behaving and how can we figure out how to teach them more effectively? But Eddington reminds us:

> *It is also a good rule not to put overmuch confidence in the observational results that are put forward until they are confirmed by theory*.[4]

While this sounds like a strange transformation of the usual, "Don't believe any theory until it's been confirmed by experiment," it has a solid grain of truth. Until we have a way of making sense of, organizing, and understanding our observations, we may well misunderstand what those observations have to tell us.

The issue of how to build a coherent mental picture (theory) of what happens in a student and a classroom is the missing leg of our three-legged stool. While many educational theories exist, they are often narrow prescriptions that offer heuristics for improving instruction. We need something that can providing a structure for interpreting observations, for developing and testing models that can grow and accumulate knowledge scientifically, and can guide the creation of appropriate methodologies.

Part of the challenge in building educational theory is that learning is a human behavior and human behavior is extremely complex. At this stage, we cannot expect to have anything like a complete theory. But it is clear that whatever we do, we have to consider how thinking works (cognition) and how individuals interact with the context and cultures around them (situational and sociocultural interactions). If we don't yet know enough about how these work, what can we do? Just as we have done in many areas of science, we create a theoretical framework that allows us to build descriptive models and that can evolve and change as we learn more.[5,6,7]

When we teach physics, at least before graduate school, we focus our instruction on well-developed theories. These come with well-established models – touchstone examples – that are often viewed as integral parts of the theory, since they show how the theory works and is applied. When we are trying to establish a new theoretical framework, it is useful to separate the "bones of the framework" – the basic assumptions of what kind of things we are talking about and their nature (the *ontology* of the theory) – from the specific models and examples that "flesh out the framework". In order to clarify what I mean by a theoretical framework, let me turn to examples that we can pull apart to find the frameworks: Newtonian mechanics, quantum field theory, and evolution.

## Framework 1: Newtonian Mechanics

Newtonian mechanics is the theoretical framework that was developed to describe motion on the human level, but it turns out to work over a huge range of phenomena, from molecules to galaxies. Newton's three laws provide the framework, implicitly establishing an ontology.

The first law

> *An object at rest tends to remain at rest; an object in motion tends to maintain its velocity unless acted upon by unbalanced forces*.

tells us that mechanics is about *objects* and *interactions with other objects* (forces). It also tells us how to describe an object: specify its position and its velocity. These variables establish the dynamic parameter space in which the object is situated.

The second law

$$\vec{a}_A = \vec{F}_A^{net} / m_A$$

reaffirms that we are talking about objects (the subscript "A"), tells us *how* the forces the object feels are to be combined (the vector symbol and the superscript "net"), and what characteristic of the object we need to know in order to understand how the interactions change the velocity (the object's mass). The third law

$$\vec{F}_{A \to B} = -\vec{F}_{B \to A}$$

tells us that forces that act between objects are *interactions* and constrained (by consistency conditions) to be mutual.

This is more complicated than it looks. In the Newtonian framework, each object gets a Newton's second law equation of its own. The connection between objects is made through the forces. In principle, we have to solve Eqs. (1.1) for all the objects we are considering at the same time. In practice, we often make easier-to-handle models. For example we might consider the earth as effectively infinite in extent ("flat-earth gravity") when we consider falling bodies. Or we may consider the sun as being fixed in space when we model the motion of the solar system.

Every example we consider in Newtonian mechanics is a model in the Newtonian framework — an approximation in which many real-world factors have been ignored. Since the framework is about objects and their positions as a function of time, it's natural and convenient to start with models that restrict the number of variables we have to deal with – such as ones that contain point masses or rigid bodies.

No one has ever seen a point mass or a rigid body, but they are useful if we carefully limit the situations in which we apply them. No one has ever seen a spring that can be either stretched or compressed by arbitrary amounts and that responds with anything remotely like a linear relation – except when the deformation is severely restricted to a limited range, but Hooke's-law models serve as excellent starting points for making sense of any kind of oscillation – and even as starting points for models in other frameworks, such as the description of the electromagnetic field in quantum field theory. The Newtonian framework is not the models we use it for. It is far more reliable than any specific model. When our models fail, we first attempt to modify the model, not the framework.[8]

## Framework 2: Quantum Field Theory

When we get down to sub-atomic and sub-nuclear phenomena we need a different framework: quantum field theory (QFT).[9] Because of the small scale and the high speeds involved in sub-nuclear reactions, quantum mechanics has to be combined with special relativity in order to guarantee that all signals propagate at the speed of light or slower. As a result, we can include no rigid objects of finite size. Everything has to be described in terms of point masses and interactions that propagate at a finite speed.[10]

The ontological structures established by this theoretical framework are quantum fields and their states, something very different from localized objects. These fields are functions of space and time that describe the probability of finding a particle at a particular point at a particular time. The "kind of thing" they are is specified by how they change when we change the perspectives and arbitrary choices we have. (This requires that we pay attention to group theory. How the fields change specifies their internal angular momentum, or spin, and their quantum numbers.) Then there are interactions – scalar functions built by combining fields at a point to yield an interaction term in a Hamiltonian or Lagrangian. So the "objects" in this framework are fields and the "interactions" are scalar products of fields.

The QFT framework leads to model building in very different ways from the Newtonian. We choose mathematical transformation groups, fields, and interactions to mount an overall "theory" (really a model). One way to approach classes of problems in QFT is to generate an infinite series using perturbation theory in which each term is represented by a Feynman diagram. We then make models of a particular phenomenon by choosing a subset of the diagrams that are assumed to be most relevant.

Note how the language and indeed the entire way of speaking differs for QFT as compared to Newtonian mechanics. Forces, fundamental in Newtonian mechanics, don't even appear in QFT. QFT and Newtonian mechanics provide different frameworks for describing different classes of physical phenomena.

## Framework 3: Evolution

Other fields of science and math also establish and work within theoretical frameworks. Evolution is a theoretical framework that biologists use to explore questions about species, their history and genetic structure, and their relationships. It can be used for an extremely wide range of situations and questions ranging from the transformations of viruses on times scales of months or years to the transformation and development of species over millions of years. It provides insight into structure and function of biological organisms ranging from the subcellular to the ecological.

Evolution is often described as a "theory", but I like to describe it as a theoretical framework. It tells us what to pay attention to – heredity (genotype), variation, organisms and their structure (phenotype), and natural selection. This encourages biologists to define species (populations that share and exchange genetic material) and to look at interactions between populations, such as predation, symbiosis, parasitism, etc. But for any particular biological question, biologists will build models within this theoretical framework, paying attention to what is most likely to be important for the description of the particular phenomenon – and ignoring everything else. For example, a model explanation of the speciation of Darwin's finches focuses on beak shape and how beak morphology can lead to increased success in particular ecological niches.[11]

These three examples give some indication of the value and power of theoretical frameworks in guiding scientific understanding and progress.

## III. DO WE REALLY NEED A THEORETICAL FRAMEWORK FOR PER/DBER?

Despite the clear value of theoretical frameworks in scientific research, DBER scientists are often reluctant to situate their work within a theoretical framework. I am sometimes told by my PER colleagues, "Why do we need a theory? We know physics and have experience learning and teaching it. I'm happy relying on that."

As scientists and science teachers, we may have experience practicing, teaching, and learning our science, but we still should be cautious in relying on our spontaneous interpretations about thinking and learning from everyday experience – even from our professional experience. After all, from three decades of PER we have learned that despite having nearly 20 years living in the physical world and experiencing motion and forces, students' spontaneous or "folk" models[12] of physics can lead them dramatically astray in trying to make sense of the physics they are learning.[3]

In the same way, as instructors, our "folk models" about how people think and learn can lead us dramatically astray in trying to make sense of how our students are responding in our classes. A theoretical framework, while not giving all the answers, can guide our thinking, alert us to things we might otherwise miss, and help us reinterpret our folk models carefully and consistently. Moreover, a theoretical framework can help us formulate the research questions we need to explore to improve both our understanding of student responses and our instruction. A theoretical framework helps us explicate our tacit (often unnoticed) assumptions, test them, and modify them when needed.

In this paper, I outline the theoretical framework of Resources and discuss one detailed example of how our spontaneous models of thinking and learning may miss what our students are doing. When our students miss a question whose answer they should know or be able to work out as a result of what we have taught them, we typically assume that that they have failed to learn or understand what we have taught them. If many students do the same thing, or if some students repeat the same mistake despite repeated instruction, we may say they bring a *misconception* into our class and expect that they will need a strong remedial effort to fix it.

Sometimes this is the case. However, the two-level version of the resources theoretical framework I describe below, built on solid results from psychology,[13,14] neuroscience,[15,16] socio-linguistics,[17] and anthropology,[18,19] suggests that errors of this type can have other causes than failures of knowledge. As we shall see, mental responses are highly dynamic, responding not only to what knowledge is available, but to in-the-moment readings of what knowledge is relevant at that instant. So errors, even reliably reproducible ones, may occur not only because of lack of knowledge but also through failures in the moment of *control structures*: mismatches of situations and expectations about the task that result in students *failing to access knowledge they actually have*. In order to understand how this works, we need to establish a few basic psychological results.

## IV. GROUNDING OUR FRAMEWORK ON BASIC PSYCHOLOGY

The community with which I identify talks about a *Knowledge in Pieces* (KiP)[20] or a *Resources*[21] framework for educational theory. I prefer the latter term since it seems more general and less constraining, allowing the combining of mental "pieces" into reasonably stable structures. As of this writing there are many papers describing work within this framework. Check out one of the on-line bibliographies for a long list.[22]

What's the appropriate level of description for a system as complex as a science classroom filled with functioning human brains? Despite my appreciation of the value of reductionism in physics, at this point I do not expect us to find a micro-to-macro fundamental theory that begins with the basic elements of thought – neurons and chemicals in the brain. Rather, we are looking for a macro-level description to guide phenomenological modeling; something in the spirit of Newtonian mechanics of point masses and rigid bodies (that doesn't require a discussion of atoms) or Kirchhoff's principles of electric currents, potentials, and resistance (that doesn't talk explicitly about electrons).[23]

Because we are building a theoretical framework for learning, our foundational knowledge naturally comes from psychology and the social sciences. While many physical scientists are still leery of the behavioral sciences, these sciences have made tremendous progress in the past few decades. While much uncertainty still remains, psychology, neuroscience, social science, linguistics, and sociology are beginning to converge on a model of how the human mind works.

We are going to use some of the fundamental ideas from these disciplines as foothold principles around which to construct our theoretical framework. My metaphor here is climbing a cliff face or climbing wall. By *footholds* I mean basic principles that we can use to organize our thinking and move onward and upward. This phrase implicitly carries a weaker claim than "laws" or "principles" and includes the implication that we may be willing to retreat and modify them – choose new footholds – if the ones we have don't take us where we need to go.[24]

We do have to be careful in taking results from the behavioral sciences. In much of their research (as in the physical sciences) there is considerable interest in fundamental mechanism. As a result, many studies are what we in physics would call "zero friction" experiments, where the context has been dramatically constrained and simplified to illuminate some basic phenomenon. We know there is great value to such knowledge, but we also know that the real world is often dominated by friction, so simple experimental results need to be considered in a realistic context, and their relevance may change dramatically as a result.

As in building physical frameworks, we want to have a few basic and powerful principles that let us do a lot. But can we get away with this? The brain is an amazingly complex and flexible device, capable of creating art, science, and culture. In our desire to have something tractable and easy to work with, we have to be careful not to create something *too* simple that does not take into account the full possibilities of the brain's dynamics and creativity. On the other hand, we don't want to get lost in trying to model the very fluid and dynamic functionality that appears to be the workings of the active, thinking brain.[25]

An analogy that gives me hope and the courage to move ahead is the atomic shell model. Consider, for example, a Calcium atom. It has a nucleus and twenty electrons crowded into a spherical volume with diameter less than a nanometer. Any pair of those electrons getting as close to each other as half a nanometer contributes a repulsive energy of ~3 eV! This is a huge number on the scale of atomic excitations. Getting two electrons closer than that costs proportionately more energy. One might expect that the most appropriate model of an atom would have to be dominated by electron-electron correlations – creating a structure where the electrons move to avoid each other as much as possible.

This is of course NOT what we do. We consider a model (the atomic shell model) where we treat one electron at a time, "smearing out" all the others and describing the motion of each electron in a *mean field* – the average field of the other smeared electrons. The result has each electron being independent of all the others in a quantum state called an *atomic orbital*. We mostly ignore pairwise electron-electron correlations. All of chemistry is based on this unlikely starting point! We later learn that when expressed in terms of these orbitals, the Pauli Principle drives the modifications to the atom's wave function due to electron-electron interactions up to high energy and, as a result, down to short range justifying the model after the fact.

The brain also has lots of strong stuff going on. Our brain allows us to store huge amounts of knowledge in our *long-term memory*.[26] Interactions between those memories are critical, since our long-term knowledge store is not well indexed – it's not even time stamped or location marked. We access various bits of memory and identify their relation to other elements (such as dates and places) through chains of associations. We build up local associational structures such as schemas, mental models, and blends. With so many bits, pieces, and interactions, how can we possibly hope to make sense of what's going on as students try to bring the large and complex body of scientific knowledge into their already complex existing structure of memory?

I propose that the functioning of the brain also has a structure that suppresses much of its apparent complexity. When we use and manipulate our knowledge, everything has to go through a very small structure known as *working memory* – what you hold in your mind and can manipulate explicitly. Before getting into details, let's explore some of the critical phenomena in your own brain.

## Experiment 1: Limited working memory

Our first experiment demonstrates the critical result that our brains have difficulty in managing tasks of too high a complexity at one time. For this task you will need a partner. Have your partner read you the following strings of numbers and after each one, try to quickly say them back in reverse order. So if your partner says "123" you respond "321". Now try it with the following number strings:[27]

- 123
- 4629
- 38271
- 539264
- 9026718
- 43917682
- 579318647

Get the idea? It gets harder and harder and above a certain string length it's impossible. (You can develop techniques to do this task, but the experiments I present are designed to show the limitations of the untrained brain.) George Miller[28] proposed the limit of "7 ± 2" on processing capacity more than 50 years ago. These observations are the basis of the important psychological construct of *working memory* – the part of your brain that you use to think with and that manipulates bits of the large store of knowledge your long-term memory contains.[13]

To see that this result has implications beyond this trivial example, take a look at A. H. Johnstone's Brasted Lecture.[29] In it, he reports on a chemistry exam on the topic of the mole (Avogadro's number of molecules) set by the Scottish examination board and given to 22,000 sixteen-year-old students. When considered as a function of the sum of (1) the pieces of information given in the question, plus (2) the additional pieces to be recalled, plus (3) the number of processing steps required, student success shows a dramatic drop off at six pieces of information, consistent with Miller's hypothesis.

Of course we all know that we can handle much more complex knowledge than 7 ± 2 items. One approach the brain uses is *compilation* or *chunking*: creating strong associations among clusters of related knowledge allows us to manipulate blocks of knowledge. Another way the brain extends the reach of working memory is using external objects as components of our working cognition, components like equations written on a whiteboard or menus in a computer program. A third technique – and one that can cause trouble in our classrooms – is memory reconstruction, as shown in our next experiment.

## Experiment 2: Reconstructive recall

In our second experiment, consider the 24 words given in the list shown in Fig. 1.[30] Look at them for one minute and try to memorize as many as possible. Don't do any-thing special or organized*: just look at the words and try to remember as many as you can. After one minute, look away and try to write down as many as you can recall.

| Thread | Sewing | Bed | Blanket |
|---|---|---|---|
| Thimble | Cloth | Rest | Doze |
| Pin | Sharp | Awake | Slumber |
| Haystack | Injection | Tired | Snore |
| Eye | Point | Dream | Nap |
| Knitting | Syringe | Snooze | Yawn |

*Fig. 1: The list of words to try to memorize for experiment 2.*

Now look at your list. Check endnote [31] to see if you had either of the two test words on your list. When I give this task to my class, typically more than half of the students put one or both of the test words on their list and are shocked to discover that they aren't there. They were sure they remembered seeing them!

This illustrates a critical principle of memory: *Memory is not veridical*. It's not an accurate recording, but rather is "reconstructed" from stored bits and pieces and plausible stock items. A lot of psychological research supports this, going back to 1932.[32,33] A readable modern theoretical interpretation (with support from neuroscience experiments) is summarized in Buckner & Carroll.[34]

### Experiment 3: Selective attention

For our final experiment, you need an Internet connection. In this task, a group of six students (shown in Fig. 2) serve as two teams, one with white shirts and one with black. Each team has a basketball and during the short video they move around quickly, passing their ball among members of their own team. Your task is to see how well you can concentrate by counting the number of passes among the members of the white-shirted team. You have to pay careful attention, since things happen fast!

Go to the link
http://www.youtube.com/watch?v=IGQmdoK_ZfY and maximize the screen without reading any of the text there (or below) until you are done.

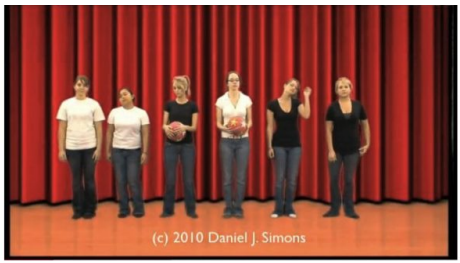

*Fig. 2: Daniel Simon's concentration task.*

Many people manage to count the number of passes successfully, but fail to see the dramatic events and changes that take place during the clip that are identified at the end. This surprising phenomenon is called *inattentional blindness* –when you are paying attention to something you think is important, you may fail to notice other important things. The fact that we pay attention to some things and ignore others (often without conscious decision) is called *selective attention*.[35,36]

This is the psychological core of the phenomenon that I refer to as *framing*, which is the heart of the phenomena about teaching and learning that I want to emphasize here. Let's begin by establishing the bare bones of the Resources Framework and showing how it follows from the basic psychological principles established by our three experiments. As our experiments show, our brains can get overloaded from the complexity of a task, fail to reconstruct our memories correctly, and even misinterpret our direct perceptions. Fortunately, the brain has lots of tools to help us do better.

Two useful and effective structures that help us access and use our knowledge are local and global associations. By *local associations,* I mean direct and immediate links – the kind of quick automatic connections that come up in free association or those kinds of connections typically drawn in a concept map.[37] Clusters of local associations can form tightly bound chunks of knowledge that I refer to as *compiled,*[7] or looser clusters of associations referred to in many areas of behavioral science as *mental models,*[38] *schemas,*[39] or *coordination classes*.[40] These structures and other associated local patterns of association are important in understanding teaching and learning.[41] Compiled or tight associations are things you always associate and find hard to break apart. For people who are good readers, it's difficult to see the letters "CAT" and not associate to one of the many meanings of the word. Looser patterns are ones where you may or may not come up with a well-known association depending on the context.

There are lots of things to say about local associations and a large interesting literature about them. But in this paper I want to focus on equally important but less often discussed structures: *global associations* that act on and control our local associations. By this I mean our use of our knowledge about situations (especially social environments) that we use as filters to restrict our search space in long-term memory. We might call them *expectations*. Let's see how what we did in the last section can help us understand how this works.

Although huge amounts of knowledge are stored in long-term memory, access to it is restricted by the psychological ideas demonstrated in the experiments in section 4:

---

* Many of you are able to construct methods that allow you to remember all these words. This kind of thinking is what we are trying to teach our students to do! To illustrate naïve student thinking, try to do this task without using any highly developed learning skills.

- Working memory – what you can hold and manipulate in your mind at any one instant – is limited.
- Memory is often not just direct recall but is reconstructive and dynamic.
- Selective attention matters – a lot.

Of course these few results don't tell the whole story. (For more discussion and lots more references, see [6] and [7].) But they can help us begin to set up some foothold ideas that have powerful implications for how we think about educational issues. I start with a brief outline of the fundamental ideas about how the brain works – cognition. I then discuss how broader knowledge plays a role in organizing thinking – especially socio-cultural knowledge. Finally, I tie them together through the process by which the brain uses socio-cultural knowledge to restrict search spaces in cognition – framing.

## The cognitive structure

To figure out how the brain functions dynamically, we need a somewhat mechanistic picture. I've outlined a model in Fig. 3. Let's imagine that the brain is presented with a straightforward set of data: the perceptual signals associated with holding a cup of Turkish coffee. These include a variety of sensations: (1) *visual* – a pattern of signals arriving on the retinas of your eyes, (2) *haptic* – the sense of touch including the feel, texture, and weight of the cup in your hand, (3) *olfactory* – the smell of the coffee, and (4) *memory* – your knowledge of the cup, including how it tastes, what the effect of the coffee might be on you, and your social knowledge about how and when to drink it – and when to *stop* drinking it so you don't get a mouthful of grounds.

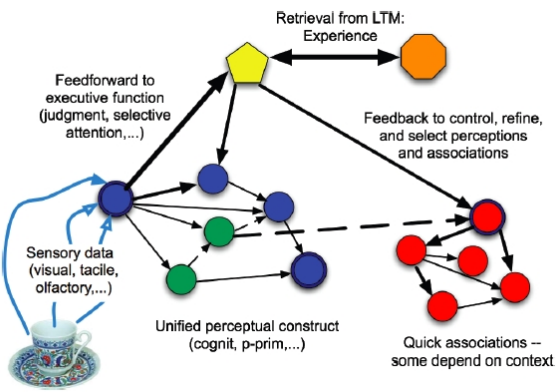

*Fig. 3: Dynamical structure of the brain's response to input data.*

The first step in the way the brain appears to work is that basic sensory data is processed to create a coherent perceptual construct and activate associational knowledge. While it is doing this, it sends signals to the judgment and decision-making part of the brain, the pre-frontal cortex. This part of the brain accesses information from long-term memory to decide what to do with the data. This is where knowledge about the way the social world works is brought in.[42] Selective attention (such as in experiment 3) happens here and other perceptions and associations (such as in experiment 2) can now be linked to the original percept.[15,16,43]

This model sets us up the structure for the phenomenology that I will use to analyze and describe the role of context and expectations: associations and control of those associations. Here are two foothold ideas.

- *Associations* – Activity in the brain consists of turning on some bit of knowledge (*activation*). This bit of knowledge links to others and sends out signals that tend to activate (or inhibit the activation of) other bits. Activation of one cluster may induce activation of other clusters leading to interpretation and meaning making.[44]

- *Control* – The brain takes data from immediate situations, uses it to activate generalized situational memory, which then feeds back to the system to control selective attention. The feedforward and feedback of signals to and from the prefrontal cortex and long-term memory activate or inhibit the activation of associational clusters.

The control level is where students' assumptions, expectations, and culture draw on their broad knowledge of appropriate behaviors to affect what they do in our classrooms. To understand how to talk about this, let's consider how the behavior of an individual is imbedded in a socio-cultural environment and how this environment affects behavior.

## The socio-cultural structure

The behavior of any human being is immensely complex. It can be analyzed at multiple scales, ranging from the very small (how many neurons are being activated) to the very large, both in space and time (how does it depend on the presence of highly structured modern technology or the modern nation state). It responds to the individual's knowledge of the human social world which comes from many sources and scales.

I show one way of thinking of this in Fig. 4. I borrow a metaphor that has been used by complexity theorists as a metaphor for resolution or "grain size": the staircase. When looking at something while standing on the bottom (and looking down) you can see all the local detail on the ground in front of you. The higher up you are, the less detail you see – but you are better able to discern broader emergent patterns. In complexity theory, moving up the staircase is *reductionism* – explaining things at a larger grain size in terms of finer-grained structures. Moving down the staircase is *emergence* – organizing knowledge at smaller scales in terms of larger scaled structures that emerge from coherences and collective effects (and that can be hard to understand from a reductionist perspective). I'll refer to it as the *grain-size staircase*.

On the finest level, we see neurons and their functioning – the fundamental matter of which behavior is made up. When we move up a step, we are led ask ourselves about the basic psychological mechanisms of behavior – what they are and how they develop. Another step up takes us to basic behavioral phenomenology – what individuals know about the physical world and how they interact with it.

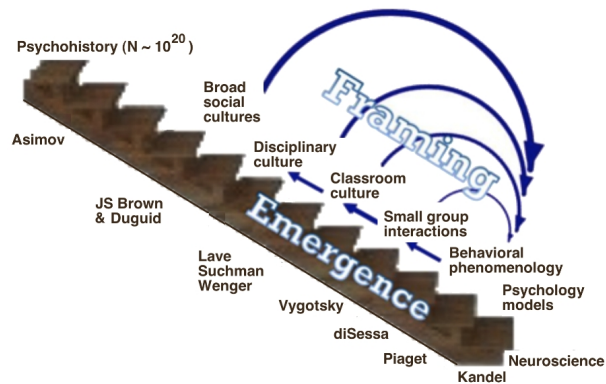

*Fig. 4: The cognitive/socio-cultural grain-size staircase.*

The next step moves beyond individuals and places them in the context of a small group. Beyond that, we consider individuals' relations to the broader local culture of the environment – their knowledge and experience with classrooms and school and their understanding of appropriate behavior in that context. The classroom itself then gets imbedded in multiple cultures – the culture of the discipline being taught[45] and the way schooling is imbedded in the broader culture of the locale[46] – how schooling tends to be viewed by other individuals in the society, how it relates to employment opportunities, how one's position as a member of various subgroups in society affects one's behavior, and so on. Power relations, stereotypes, and other important factors come in at this level.

At another step up, we might begin considering the behavior of groups of individuals whose function has to be viewed as a group. A software development corporation may have coherent capabilities that no individual possesses;[47] a battleship may know how to navigate, but no single individual in the navigation team may have that knowledge.[48]

The same idea holds for science. Scientific knowledge is an emergent community consensus that arises from the knowledge of many individuals. No single individual, no matter how brilliant, has knowledge that is identical to this community knowledge.[49]

Each level is emergent from the level below it. The critical behaviors seen at a given level are a result of structures at lower levels, but how they emerge from those structures may be difficult to analyze. Note that the linear model of the staircase is somewhat misleading. Phenomena that are organized at multiple levels may affect each other and interact. The higher levels, which can be seen as emergent from the lower, feedback and constrain phenomena viewed at the lower levels. The metaphor only is meant to specify that some phenomena are most naturally considered at different scales; it is not meant to constrain how those scales interact with each other.

The important part of this staircase analysis for us as science educators is that the behavior of the individual student in our classroom is affected – often strongly – by the knowledge they bring to the classroom. And their knowledge and perception of <u>all</u> the upper levels of the staircase can play a critical role, serving as control structures for what behaviors they engage in and what they avoid. I show this in the figure as arrows looping back to the basic behavioral level.

In order to have a language to talk about how this works, I adapt the process known as *framing* from anthropology[50] and sociolinguistics.[51]

## Framing: The interaction of the cognitive and the cultural

Socio-cultural effects on the classroom have been studied extensively for many decades, but often a critical point is not made explicit. It's not just the socio-cultural environment that matters: rather, *it's a student's perception of the socio-cultural environment that affects that student's behavior.*[52]

This requires us to not simply look at the environment and interpret it through our own perceptions, but to consider what socio-cultural knowledge the student brings to our classroom and how the student uses that knowledge. As in experiment 3, if our students fail to perceive what we have set up for them or asked them to do, it might as well not be there.

The anthropologist Erving Goffman studied how people interpret and respond to the social environments they find themselves in from moment to moment.[18] He suggested that people are continually asking themselves the question (though not necessarily consciously), "*What's going on here?*" The answer to that question controls (again, not necessarily consciously) what behaviors the individual activates. Goffman referred to the process of answering that question by drawing on experiences stored in long-term memory as *framing*. The concept has been further developed in sociolinguistics[17] and in other fields as well.[53] For a detailed discussion of how it applies in physics education, see Hammer, et al.[54]

Framing is what you did when you focused your attention on the passes in experiment 3 and as a result wound up not seeing other elements that were interesting and possibly important. In that case, in my instructions, I encouraged you to frame the task as a concentration one, which encouraged you to ignore (or even suppress) everything else that might be happening. Similarly, problems occur in a classroom when students bring in their own expectations that may result in their ignoring messages that you think you are explicitly sending – expectations like, "I know how a science class works. I don't have to read all these pre-class handouts."

## V. HOW THIS WORKS IN PRACTICE: CLASSROOM EXAMPLES OF EPISTEMOLOGICAL FRAMING

A number of broad surveys – for example the MPEX,[55] C-LASS,[56] CHEMX,[57] and MBEX[58] – probe students' attitudes and epistemological expectations about what and how they will learn in an introductory science class. These consider not just students' general attitudes about science, but their epistemology – what they think they know about the nature of scientific knowledge, but also their *functional* epistemology – what they think they have to do to learn that knowledge.

Pre-post results are rather depressing. Students enter the class with expectations that are somewhat in accord with their instructor's (about 2/3 of the time), but they leave it having either remained the same or deteriorated. These surveys provide a strong indication that we have a problem that goes beyond the well-documented conceptual learning difficulties that students show.

To better understand both how students' epistemological expectations play out in an actual classroom environment and how we might affect these dimensions of learning, we need a closer look at how students function in a classroom than can be provided by simple pre-post testing. We get this through videotaping students – both in semi-structured interviews[59] and in classrooms where much of the learning takes place in a group-learning environment. In many of the latter situations, students interact with their peers in the absence of an instructor, for example, when working in groups to solve homework problems.

These videotapes can give us insight into how control structures actually function to foster or hinder deep learning. The results discussed in this section come from hundreds of hours of observations with physics students at levels from introductory college up to graduate school.

When students are put in a situation in which they have to construct some knowledge – answer a question, solve a problem, analyze a text or experiment – they make a quick and dynamic decision (again, not necessarily consciously) about "what is going on". They decide how to restrict their *search space* in their long term memory – what knowledge they have that might be relevant to bring to bear and how to approach what they need to do. I refer to this control structure as *epistemological framing* — the process that generates each individual's answer to the questions:

> *What is the nature of the knowledge*
> *we are learning in this class*
> *and what do I have to do to learn it?*

Epistemological framing is the process of choosing different ways and tools of knowing (*epistemological resources*) for dealing with learning situations.[60,61] This choice of metaphor ("framing") emphasizes the *process* aspect of the phenomenon. We can also choose to talk about an *epistemological stance*, choosing a metaphor that emphasizes the *functional state* produced by the framing process.[62]

Epistemological framing restricts the conceptual resources, associational structures, and epistemological resources[63] that students access in a particular context. Some epistemological stances that we have observed are discussed in stories in which students are led astray by four identifiable epistemological framings.

- *One-step thinking* – "The answer is obvious. I don't have to worry about context or coherence."

- *P-priming* – "The answer is obvious. I don't have to worry about how it works."

- *Rote reasoning* – "I know the process to generate this answer. I don't have to think about meaning."

- *Disciplinary siloing* – "Since this is a physics course, I don't have to bring in any knowledge from chemistry."

While these are sometimes perfectly appropriate, in some situation they lead students into trouble.

## Framing 1: One-step thinking

One of the most common inappropriate epistemological framings that I encounter among students in my introductory physics classes is the idea that, "I should know every answer right away through direct recall." This often leads them to answer on the basis of a single remembered association without thinking through what the question means or relating their answer to other things they know. I call this *one-step thinking*. An example is given in Fig. 5. I gave these problems in an algebra-based physics class using clickers at the beginning of the second term as a part of a discussion about how to learn to think in the class.[24]

The first problem, labeled A, asks, "Given that a light and heavy ball fall to the ground at the same time, which is being pulled harder by the force of gravity?" The second rotates the motion from vertical to horizontal, reverses the direction of the implication, and asks, "If you are pulling a light and heavy object with the same force, which will speed up faster (ignoring friction)?"

Although it seems obvious that the force of gravity on a heavier object is bigger than on a light object (after all, that *is* what "heavier" means), half of the students answered the first question saying that creating equal speeds required equal pulls on different masses. But more than 80% realized that creating equal speeds for a heavy and light mass requires more force on the heavier. (It doesn't matter which question I ask first. I get about the same result either way.)

Discussion makes it clear that most of the students who answer the falling body question wrong seem to be framing the task as a "physics recall" problem and as a result are not accessing their intuitive sense of how to make heavy and light objects move. They see "gravity" in the first question and remember from Physics 1, "Oh yeah. We had a funny result about falling and the implication was that gravity was the same." They reconstructed a memory and did not include in their epistemological framing the need to check their memory against their intuition.

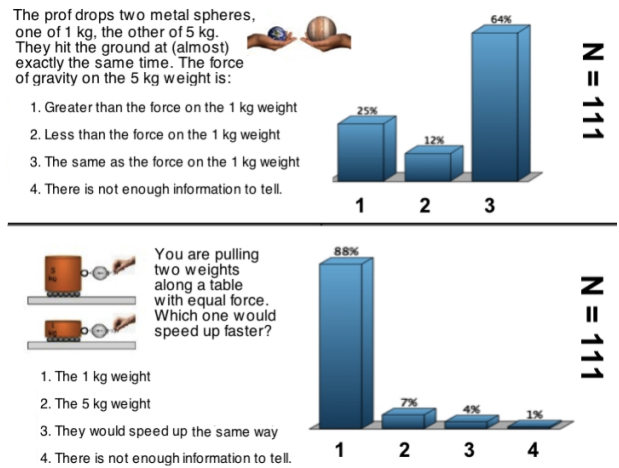

*Fig. 5: Student responses to nearly equivalent questions.*

Their error is not that they don't know that it takes more force to accelerate a heavy object than a light one. Their error is they don't bring that knowledge to bear in this particular context. The failure to demand reconciliation of the physics they are learning with their everyday experience is a familiar problem to instructors of introductory physics and raises serious barriers to our students developing good physical intuitions.

The problem is not only that students do one-step thinking in response to a question and get the wrong answer.

Sometimes the wrong epistemological framing leads them to shut down and not even try.

In my exams, I often give questions that students have not seen before, but which they can solve if they know the foothold principles we have discussed and how to apply them. At the beginning of the term, I tell them that they will have to think on exams, but many ignore my statement – in part because they are not sure what it means. After the first exam in the second semester, a student who was new to my class came in to see me because she was upset at her results. She had barely earned a C and was used to straight A's. She wanted help in understanding what she was doing wrong. I began by asking her to show me what she had done on the first problem – a 5-part short answer problem about forces among three charges in a line.[†] She had missed every one. She sighed and said, "I didn't know any of these answers so I just guessed." I responded, "Well, I don't know what you learned last term. On parts B, C, and D it asks for a net force. How did you learn to approach that?" She answered, "We did free-body diagrams." I asked her to try that technique. She proceeded to use it successfully without any help from me and went on to solve every part of the problem quickly and effectively. "Oh!" she said. "I'm supposed to figure them out!" From that point on, she earned straight A's in all her exams. Her problem was not that she didn't know the physics being tested; she had incorrectly framed the task epistemologically.

### Framing 2: P-priming

The examples in the previous section relied on students' framing a task as being answerable by direct recall from memory and, as a result, failing to access resources that they have and can use effectively. In this second framing, students treat a task as being solvable using their intuitive physical sense of how the world works (phenomenological primitives[20]), again without thinking carefully about what's really going on.

This data comes from the work of Brian Frank.[64] It takes place in a recitation of our algebra-based physics class that had been modified to place more emphasis than usual on epistemology.[24] These recitations are run as Tutorials in the University of Washington model.[65,66] Students work in groups of 3-5 facilitated by a teaching assistant trained to understand the difficulties the student typically encounter and to encourage students to explore their own ideas with each other.

The first lesson of the semester was on the concept of instantaneous velocity. We used a standard device (shown in Fig. 6). A long thin paper tape ("ticker-tape") is attached to a low friction cart (shown at the left) and run through a "tapping device" that taps a sharp point onto the tape through a piece of carbon paper at a fixed rate. The cart is allowed to accelerate slowly down a long ramp and the tapping device creates dots on the tape whose spacing indicates the cart's speed.

---

[†] http://www.physics.umd.edu/perg/abp/TPProbs/Problems/E/E67.htm

The tape is then cut into segments of six dots each. Since the cart accelerated slowly, 6 dots (representing about two-tenths of a second) look as if they are representing a constant speed. A group of 4 students receives four segment of paper tape as shown in Fig. 7.

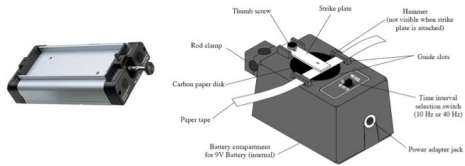

*Fig. 6: Pasco low-friction cart and ticker-tape tapper.*

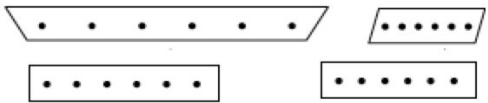

*Fig. 7: Samples of ticker tape given to the students.*

The lesson begins with the TA describing the machine that makes the tapes and how it generates them. The first question the students are asked in the lesson is: "How does the time taken to generate one of the short segments compare to the time to generate one of the long ones?"

Since the marking device taps at a fixed rate, the answer is trivial: they each have six dots, so they each took the same amount of time to make. But that's not what the students said. Here are some transcripts of our videotapes:

*Group 1*
   S1: *Obviously, it takes less time to generate the more closely spaced dots.*

*Group 2*
   S2: *(Reading) "How does the time taken…" It's shorter! (Huh!)*
   S3: *Yeah. Isn't it pretty much – The shorter ones are shorter.*

*Group 3*
   S4: *(Reading) "The time taken to generate one of the short segments…" It's shorter!*

*Group 4*
   S5: *Well it takes less time to generate a short piece of paper than it does a long one. (pause) I would assume. (pause) I don't really know how that thing works. [The last two comments are ignored by the rest of the group.]*

It's dramatic watching one group after another give the same obviously incorrect answer, confidently and without hesitation. This looks suspiciously like it's some kind of "common misconception". But the last group we quote gives a hint as to what's going on.

If we go a bit further into the videos, we find that a few questions later the lesson shifts the context. The result is that the groups bring a different approach to bear. They are asked, "Arrange the paper segments in order by speed. How do you know how to arrange them?" Here's a typical response from one of the groups:

*S1: Acceleration! It starts off going slow here, [pointing to a short segment] then faster, faster, faster [pointing to a long segment].*

*S2: No, no! Faster, then slower, slower, slower! This is slow [pointing to a long segment].*

*S1: When it gets faster it gets farther apart. That means the paper's moving faster through it. [gestures] So it's spaced out farther.*

*S2: Wait. Hold on. [gestures to TA]*

*S2: [to TA] Is the tapper changing speeds or is the paper moving through it changing speeds?*

*TA: The tapper always taps at the same speed.*

*S1 and S2 [together and pointing at each other]: Ahhh!*

They then proceed to correctly analyze what's going on. We saw this again and again. At the beginning the students gave a quick answer – longer tapes take more time, shorter ones take less. Just a few minutes later, the light dawns and they all get it right (Fig. 8).

What is changing when the students in these groups shift their behavior in response to a (very slightly) changed context? I suggest that the easiest way to describe what is happening is as *epistemological re-framing*.

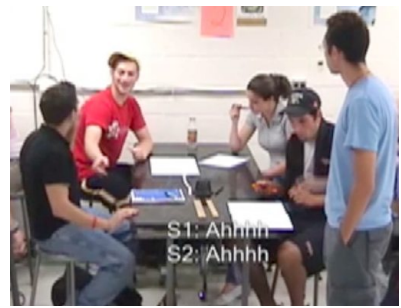

*Fig. 8: The light dawns.*

In PER we often have a tendency to refer to common errors that students bring into the classroom as "misconceptions". I don't have a problem with this, but I would like us to take a finer-grained view. I define a *misconception* as *a student error that is commonly and reliably* *activated in a given context*. Adding context encourages us to understand what is happening in detail – to consider components of the student's response, rather than using the term to close off further consideration of what the student is actually bringing to the task. Misconceptions can have structure, not just be an irreducible gallstone that needs to be excised. Misconceptions can be robust and hard to undo, but sometimes they are created on the spot and are highly context dependent. Sometimes, they contain the structures that can provide their own cure.[67]

In our example, at the start of the lesson, many students have the epistemological assumption that they will be able to generate a correct answer by simply looking at the question and drawing the most immediate and natural response – essentially by an immediate association without carefully considering the mechanism of what is happening. What they get is a *phenomenological primitive*[20] (p-prim): "more is more", which in this situation they map into "a longer tape takes a longer time". This feels right to them (partly because it is easy to generate[68]) and they move on.

The misconception in this case isn't actually a misunderstanding about the nature of velocity; it's a common framing error. The misconception is epistemological rather than conceptual: students assume that the answer can be generated directly by fast thinking without a careful consideration of the mechanism. In the last two sentences quoted, the student in group 4 expressed a framing caveat: in effect, "We might have to consider the mechanism here."

The later questions on the worksheet can't be answered without considering how the cart is moving and the mechanism creating the tapes, so a frame shift was needed. This led the students to go back and reconsider (and correct) their answer to the first question. (Of course the students were also doing "one-step thinking" and in the previous example the students were "p-primming" by choosing "same means same" in the gravity question. But in that case we are emphasizing the failure to active the epistemological resource of *coherence* and here the failure to activate the epistemological resource of *mechanism*.)

Many physics teachers are surprised when our students "miss the gorilla in the classroom" and assume they didn't need to think about the mechanism of what's happening – especially since the lesson begins with the TA describing the mechanism just a minute or two earlier! But selective attention can cause students not only to focus on particular aspects of a task but also to ignore other aspects that their instructors might consider natural and critical.

## Framing 3: Rote reasoning

Our first two framings involved students not thinking much – coming up with answers too quickly or not bringing to bear things they knew well. But framing errors don't only occur by students failing to think enough. Sometimes they reason long and hard – but still fail to bring in relevant things they know. Two examples are found in the work of Tuminaro[69,70] and Bing.[71,72,73]

Tuminaro videotaped students in algebra-based physics working on problems in groups. In one example,[70] he watched a student trying to solve an estimation problem in the section on fluids in the last week of the first semester. The problem is shown in Fig. 9.

The student thought she knew what to do. She looked for an equation for pressure and by failing to distinguish pressure from *difference in pressure,* brought up the wrong one: $pV = nRT$ (instead of $p = p_0 + \rho g d$). She then tried to see which variables she knew the value of and which needed to be calculated. She wound up making some very bizarre statements – such as insisting her dorm room should be considered to have a volume of 1 m$^3$. She failed to identify this as an estimation problem in which you are expected to quantify your personal experience. She felt that any number you needed *had* to be given somewhere in the problem. The only visible volume was in the density – 1 kg per 1 m$^3$. (This despite the fact that the class had done an estimation problem on every previous homework assignment.) Her epistemological stance for this task was what Tuminaro called *recursive plug-and-chug* – an explicit problem-solving process that she had learned in high school. It leads to successful calculations in many situations, but it does not typically include sense-making or deciding whether an equation is relevant or meaningful.

In a second example, Bing studied students working in groups in upper-division physics classes. In one example,[72] he watched a group of half a dozen students solving a quantum mechanics problem from Griffiths.[74] As part of studying the role of symmetry in two-particle wave functions, they had to calculate the expectation value of $<x^2>$ in an excited state of the one-dimensional particle in a box. One student wrote down the integral

$$A^2 \int_{-\infty}^{+\infty} x^2 \sin^2\left(\frac{n\pi x}{L}\right) dx \ .$$

This is the correct integrand, but with the wrong limits. She proceeded to spend more than 15 minutes using a variety of tools to try to construct the integral – Mathematica™, analytic integration by parts, an integration tool on a TI calculator. Throughout, she showed good mathematical thinking and convinced herself (correctly) that the integral diverged. A second student was growing increasingly uncomfortable. "We've done that integral so many times... (starts paging through in the book)." Having convinced herself with extensive reasoning, she challenged him to find it. After some time, a third student said, "Hey! It's not negative infinity to infinity…we just have to integrate it over the square well since it's the infinite square well." Everyone immediately agreed and they quickly found the correct solution. Although our first student clearly excelled at mathematical reasoning, by framing the problem as *only* a mathematical one, not one requiring the blending of physical and mathematical knowledge, she wound up on the wrong track for an extended period of time. Once she was cued to realize the physical fact she already knew, she was able to import it into the math and quickly do the integral correctly.

In our first three groups of framing examples, students failed to answer a question correctly, not because they did not possess the needed tools or understand the critical concepts; they just didn't see that they needed to access many of the resources they possessed. These weren't conceptual difficulties; they were selective attention problems – epistemological framing.

A final epistemological framing shows that higher steps in the staircase are relevant as well. This example involves the use (or not) of cross-disciplinary knowledge.

### Framing 4: Disciplinary siloing

My discussion of this last case will be brief because our work on the subject is fairly recent and is still shaking down. But I want to include it because it shows how broader cultural expectations (from a higher step in the staircase) can play a role. I call this particular type of framing, *disciplinary siloing*. It affects both students and instructors. In it, students (and faculty) classify particular bits of knowledge and types of reasoning as "belonging to one particular discipline" and fail (or even actively refuse) to call on these resources in the context of other disciplines. This is a big topic, so I will not cite specific examples, but only given general statements. See the references for examples.

In many universities, a majority of students taking physics are not physics majors but engineers, biologists, pre-health-care students, or something else. Recently, my research group has been working on how to transform introductory physics for biology majors into a course that holds value for them and for the instructors in their later biology classes.[75] As part of this effort, we have been holding extensive conversations with biologists,[76] interviewing both students and faculty about their views of the relations between physics and biology, and observing our students both working on physics problems in biological contexts[77] and exploring biological situations from a physics point of view.[78]

One thing that we have learned from our interdisciplinary conversations: Biology and physics faculty look at the goals and approaches towards introductory instruction in their discipline in dramatically different ways.

Here are two of the differences we have seen:[76]

- Physicists tend to want to express their knowledge — even their conceptual knowledge — in terms of equations and mathematical relationships and are accustomed to thinking using mathematical manipulations, even about qualitative issues. Biologists are much less likely to want to do this and may view such activities as valueless or even misleading.

- Physicists are fond of "toy models" — highly simplified situations that can be carefully and completely analyzed mathematically. These become "touchstone examples" that illustrate a particular (often mathematical) method or principle and serve as metaphors for modeling more complex situations. While toy models are broadly present in introductory physics courses, they also appear in the advanced research literature. Biologists much prefer to tie their discussions and analyses to real world examples. The powerful connections in biology between structure and function lead them to be highly suspicious of models where the structure appears to be "too simplified".[79]

These disciplinary epistemological stances create barriers between physics instructors and their biology students, and between physics and biology faculty trying to negotiate how to create an effective multidisciplinary curriculum. We have a large amount of data supporting these claims, including interviews with students in a biology class ("I don't like to think of biology in terms of numbers and variables…. I can't do it. It's just very unappealing to me."[80]) to Likert-scale attitude surveys of biology students in a traditional physics class reporting on the value of physics for biology (favorable results fell from 57% before the first term to 40% at the end of the first term, and to 37% at the end of the second[58]).

Although I do not have as extensive data on engineering students, I suspect disciplinary siloing occurs there as well, with engineering students being much more interested in "doing things" than with the theoretical principles that allow them to do those things. The fundamental engineering interest in *design* is rarely reflected in an introductory physics course for engineers and I expect this creates problems as well.

### The take-home message:
### It's not just concepts. Framing matters.

Epistemological framing – what students think is the kind of knowledge they are seeking and what they think they have to (or are allowed to) use to get it — often plays an important role in how they behave in our classes. If we ignore the issue of epistemological framing, we might misinterpret where a common student problem lies and have trouble creating an effective lesson — or fail to understand *why* a particular lesson is (or is not) effective.

Here's the takeaway message:

> Student responses don't simply represent activations of their stored knowledge. They are dynamically created in response to their perception of the task and what resources are appropriate to bring to bear. As a result, their behavior may have a complex structure. The (often unconscious) choices they make as to how to activate, use, and process knowledge are often determined by their social and cultural expectations (framing).

We need to keep in mind the possibility that students may not just be "wrong", "not know", "not understand", or "exhibit difficulties", but that they may rather be "doing the wrong thing".

## VI. WHAT HAVING A THEORETICAL FRAMEWORK DOES FOR MY TEACHING

The shift in perspective to include an awareness of students' control structures – their framing of their immediate context and the socio-cultural expectations they bring to bear – has had profound implications both on the way I perceive and interact with my students and on the way I carry out my education research. This deserves lots of discussion and lots of examples. But since this paper is already long enough, I will just mention a few items. (For more discussion of these and related issues, see my book about teaching[81] and my Varenna lectures.[6]) Here are four implications:

- It makes my responses to student questions more effective.
- It leads me to have more respect for students' thinking and opinions.
- It encourages me to include "meta-formative evaluations".
- It helps me generate hypotheses for research.

### It makes my responses to student questions more effective

Taking a theoretical perspective has made me rethink how I provide help to my students. Years ago, if a student came into my office hours with a question, I would give the best possible answer that I could. I was framing my task as if I were still a student, trying to give my instructor the best possible answer that I could, to show how good a student I was. This made sense at one level. I am where I am (a faculty member at a research institution) because I had been a good student. I knew how to play that role.

Now, my theoretical frame helps me see this as an inappropriate role reversal. I shouldn't have been viewing the student's question from a focus on the content of the question. Rather, I should have been trying to answer the question, "*Why can't this student, to whom I have taught this material, answer this question for herself?*" I should have been *diagnosing* her learning difficulty, not answering the content of her question. Not only did I make a framing error, but I now understand that there are many problems that might be keeping her from getting the answer on her own, and many of them may have to do with framing errors rather than lack of knowledge. Careful questioning is needed to identify what's really going on and what help the student needs. My response to the student who was unhappy about her exam grade (in the section on *One-Step Thinking*) is an example of how I go about this.

### It leads me to have more respect for students' thinking and opinions.

A second shift in my teaching has been to develop more respect for my students' responses and opinions. Increasingly, I have been giving them more latitude to answer my questions in class and have been following up on "wrong" answers more persistently. Often I am surprised by students' answers. They sometimes illuminate tacit assumptions I am making and have not been aware of.

In one example, I gave the class a clicker question[‡] to select the graph of the electrostatic potential in an infinite parallel plate capacitor. Since the E field is constant between the plates and 0 outside them, the graph is constant outside the plates and a straight line inside as shown in either of the two graphs in figure 10. Students are supposed to select one of these two from among eight possibilities. This is a standard example and I have given it dozens of times in the years I have been teaching.

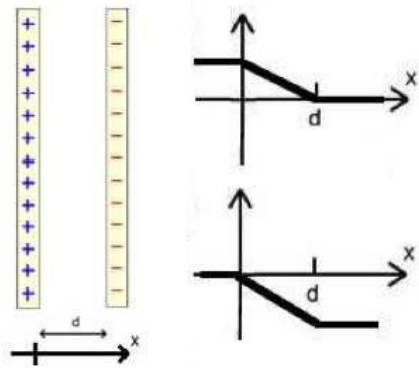

*Fig. 10: Possible graphs of the potential between two charged capacitor plates.*

Some students refused all of the answers offered and selected "none of the above." When I asked them to explain, one said, "Since all the charges are on atoms, when you get close to a charged atom in a plate it should shoot up high." Another said, "When you get far away from the plate, on either side, you have to go to zero – and it should be symmetric." I realized that what I thought was the "obvious right answer" was only right when I made a series of tacit assumptions that I had failed to explicate. This resulted in an excellent discussion about the nature of approximations and toy models in physics and their value. Since thinking about mechanism, atoms, and how things are built up from the basic physics is a part of my (epistemological) educational goals, I was extremely pleased with how it worked out.

---

[‡] http://www.physics.umd.edu/perg/abp/TPProbs/Problems/E/E42.htm

## It encourages me to include "meta-cognitive formative evaluation".

Since the theory teaches me that the mind is not always to be trusted, it encourages me to include explicit elements in my class to help students become aware of the limitations of their "natural thinking" and to encourage them to develop scientific habits of mind. I call this *meta-cognitive formative evaluation* – an evaluation that encourages students to think about their own thinking.

One way I do this is to give regular "tricky" 10-minute quizzes once a week. The quizzes are straightforward and solvable short-answer or multiple-choice problems, but are set up to lead to the wrong answer if students resort to one-step thinking is or ignore mechanism (p-priming). They often include attractive distractors like the questions in the famous concept inventories.[82,83] Often, a question will look like one that students have previously seen as a clicker question, on the homework, or in recitation – but with a change in the situation that leads to a change in the answer.

At first, students complain that these are "unfair" or "trick questions." But these quizzes each don't count for much (1% of the class grade). I stress that our goal is formative and is not about coming up with a particular answer, but rather in learning how to take a test where "thinking is required". I return the quiz and go over it in the next class, presenting the distribution of answers chosen and draw out discussions of why students picked the wrong answers. Often, these are the most valuable discussions in a week, and occasionally, have led to a deep discussion of the physics that takes up the entire rest of the class. Students often report these discussions as being extremely valuable.[§] An example of such a quiz question is given in Fig. 11. About 40% of students chose the incorrect answers [a] and [b].

> We sometimes write the symbol "g" to stand for the value 9.8 m/s$^2$. What does this stand for? Give all the answers that are correct (but you will lose points for any wrong ones you include).
>
> [a] The magnitude of the acceleration of any object feeling a gravitational force.
> [b] The gravitational force on any object near the surface of the earth.
> [c] The magnitude of the acceleration of any object in free fall.
> [d] The gravitational field strength (same as 9.8 N/kg).
> [e] None of the above.

Fig. 11: A quiz question that students found challenging.

Some of these items are insufficiently specified. Items [c] and [d] are correct if you assume you are near the surface of the earth – but that isn't mentioned. My hope is that some student will raise the issue and I will be "forced" to accept answer [e] as correct. This encourages students to think about their tests, think about their thinking, and challenge my grading. This is a much better situation for epistemological development than simply looking at it and saying, "Oh. I got it wrong. I better memorize this answer."

## It helps me generate hypotheses for research

My theoretical framework is equally valuable on my research side. When we apply our folk theories of teaching and learning – often without being explicit about it[84] – we can misread what is going on. When a lot of students make the same error, I might attribute it to a serious misconception – something that we need to help the students "unlearn". But having the two-level resources framework in my pocket leads me to hypothesize that the problem may not be what the students *know* about the physics, but about *what knowledge they have that they activate in the moment*. These two problems can lead to identical symptoms (student responses to exam questions), but they require dramatically different cures.

Even asking the students to "explain their reasoning" may not be sufficient to disambiguate these two problems. For example, if students are p-primming, they may not have a "reasoning" that they can explain. A p-prim is just an answer that "feels right" from lots of everyday experience.[20] If asked to explain their reasoning, students may generate a reason after the fact, providing not the reasoning behind their answer but a reason they think may satisfy the instructor. Disambiguation can require subtler observations, such as think-aloud protocols.

A method I have found very valuable is watching students solve problems in groups. When they disagree with each other, they articulate their reasoning, trying to convince their peers of their answer.[61] To see how this approach plays out in specific examples, see the many papers in the Resources/KiP bibliographies.[22]

## VII. CONCLUSIONS

The two-level theoretical Resources Framework that looks to understand student thinking through making sense of association structures in their knowledge and finding control structures that manage the access to those structures is not intended as a closed or "final theory".

---
§ For more meta-cognitive formative evaluation methods, see refs. [24] and [81].

Rather, it should be looked at as a "foothold idea" – a set of principles for going forward and building theory. As such, it should become an element in the partnered dance of theory and experiment, each providing support for each other, sometimes one partner leading, and sometimes the other.

## ACKNOWLEDGMENTS

The work described here has been supported in part by many NSF grants and by a grant from HHMI. I am grateful to Eric Kuo, Ayush Gupta, and Cedric Linder for comments on the manuscript.